\documentclass[fleqn,usenatbib]{mnras}

\usepackage{mathptmx}

\usepackage[T1]{fontenc}

\usepackage{graphicx}	
\usepackage{amsmath}	
\usepackage{amssymb}	
\usepackage[usenames,dvipsnames]{xcolor} 
\usepackage[normalem]{ulem} 

\newcommand{\massiveblack}{\mbox{\sc{MassiveBlack}}}
\newcommand{\bluetides}{\mbox{\sc{Bluetides}}}
\newcommand{\simba}{\mbox{\sc{Simba}}}

\newcommand{\eagle}{\mbox{\sc{Eagle}}}
\newcommand{\illustris}{\mbox{\sc{Illustris}}}
\newcommand{\illustrisTNG}{\mbox{\sc{Illustris-TNG}}}
\newcommand{\flares}{\mbox{\sc Flares}}
\newcommand{\astrid}{\mbox{\sc Astrid}}

\newcommand{\spitzer}{\mbox{\it Spitzer}}

\newcommand{\hubble}{\mbox{\it Hubble}}
\newcommand{\jwst}{\mbox{\it JWST}}

\title[FLARES V: The redshift frontier]{First Light And Reionisation Epoch Simulations (FLARES) V: The redshift frontier}

\author[Stephen M. Wilkins et al.]{Stephen M. Wilkins$^{1,2}$\thanks{E-mail: s.wilkins@sussex.ac.uk}, 
Aswin P. Vijayan$^{3,4,1}$, 
Christopher C. Lovell$^{5,1}$, 
William J. Roper$^{1}$,\newauthor  
Dimitrios Irodotou$^{6,1}$,
Joseph Caruana$^{2,7}$, 
Louise T. C. Seeyave$^{1}$, 
Jussi K. Kuusisto$^{1}$, 
Peter A. Thomas$^{1}$,\newauthor  
Shedeur A. K. Parris$^{1}$ 
\\
$^{1}$Astronomy Centre, University of Sussex, Falmer, Brighton BN1 9QH, UK\\
$^{2}$Institute of Space Sciences and Astronomy, University of Malta, Msida MSD 2080, Malta \\
$^{3}$Cosmic Dawn Center (DAWN) \\
$^{4}$DTU-Space, Technical University of Denmark, Elektrovej 327, DK-2800 Kgs. Lyngby, Denmark \\
$^{5}$Centre for Astrophysics Research, School of Physics, Astronomy $\&$ Mathematics, University of Hertfordshire, Hatfield AL10 9AB, UK\\
$^{6}$Department of Physics, University of Helsinki, Gustaf Hällströmin katu 2, FI-00014, Helsinki, Finland\\
$^{7}$Department of Physics, University of Malta, Msida MSD 2080, Malta\\
}

\date{Accepted XXX. Received YYY; in original form ZZZ}

\pubyear{2022}

\begin{document}
\label{firstpage}
\pagerange{\pageref{firstpage}--\pageref{lastpage}}
\maketitle

\begin{abstract}
The \emph{James Webb Space Telescope (\jwst)} is set to transform many areas of astronomy, one of the most exciting is the expansion of the redshift frontier to $z>10$. In its first year alone \jwst\ should discover hundreds of galaxies, dwarfing the handful currently known. To prepare for these powerful observational constraints, we use the First Light And Reionisation Epoch (\flares) simulations to predict the physical and observational properties of the $z>10$ population of galaxies accessible to \jwst. This is the first time such predictions have been made using a hydrodynamical model validated at low redshift. Our predictions at $z=10$ are broadly in agreement with current observational constraints on the far-UV luminosity function and UV continuum slope $\beta$, though the observational uncertainties are large. We note tension with recent constraints $z\sim 13$ from \citet{Harikane22} - compared to these constraints, \flares\ predicts objects with the same space density should have an order of magnitude lower luminosity, though this is mitigated slightly if dust attenuation is negligible in these systems. Our predictions suggest that in \jwst's first cycle alone, around $600$ galaxies should be identified at $z>10$, with the first small samples available at $z>13$. 
\end{abstract}




\section{Introduction}\label{sec:intro}

Identifying and characterising the first generation of galaxies is one of the core aims of modern extragalactic astronomy. Doing so will provide the essential constraints to galaxy formation models, helping us elucidate the key physics of early galaxy formation and evolution \citep{Dayal_review, Robertson_review}.


Over the last 15 years, remarkable progress has been made in studying the distant Universe, with $>1000$ candidate galaxies now identified at $z=6-10$ \citep[e.g.][]{BouwensUVLF2021}. These have come predominantly from the analysis of deep near-IR observations from Wide Field Camera 3 (WFC3) on the \emph{Hubble Space Telescope} \citep[e.g.][]{Bunker2010, Bouwens2010, McLure2010, Wilkins2011, Finkelstein2015, BouwensUVLF2021}, with a small but growing sample of bright sources, more amenable to multi-wavelength follow-up, identified from ground-based imaging \citep[e.g.][]{Bowler2015}. A number of these candidates have now been confirmed by spectroscopy, targeting Ly $\alpha$ \citep[e.g][]{Stark2010, CurtisLake2012, Caruana2014, Pentericci2014,  Schenker2014, Stark2017, Mason2019} or far-IR lines \citep[e.g][]{Knudsen2016, Pentericci2016, Hashimoto2019}. 

While the vast majority of sources are at lower-redshift, a handful of objects have now been detected at $z>10$, often combining \hubble\ and \spitzer\ observations. These include the surprisingly bright galaxy GN-z11 \citep{Oesch2016} and more recently a pair of candidates at $z\sim 13$ \citep{Harikane22}. With the successful launch of the \emph{James Webb Space Telescope} (\jwst), these sources look set to be only the first of many identified at $z>10$ \citep{Robertson_review}. \jwst\ offers the sensitivity, survey efficiency, and wavelength coverage to push well beyond the current redshift frontier. In addition, \jwst\ will provide rest-frame UV spectroscopy, allowing the confirmation of sources and the accurate measurement of many key physical properties. With the first results from \jwst\ imminent, it is essential that we have theoretical predictions in place to allow us to interpret these revolutionary observations. 

Theoretical predictions for the high-redshift Universe are available from a variety of modelling approaches. These include semi-empirical methods \citep[e.g.][]{Mason15, universemachine} and semi-analytical models \citep[e.g.][]{Clay2015, DRAGONS, YungI, Hutter21}. However, our most complete understanding comes from hydrodynamical simulations \citep[e.g][]{massiveblack, Bluetides-I,  Vogelsberger2020, astrid}, particularly those incorporating full radiative transfer (RT) \citep[e.g.][]{Renaissance, codaI, sphinx, codaII, thesan}. 

The main drawback of hydrodynamical simulations - especially those including RT - is that they are computationally expensive, limiting their volume, resolution, and/or redshift end point. This is a particular problem at very high redshift, where the source density is so low that simulations comparable to, and ideally much larger, observational surveys are essential to yield useful statistical predictions. Indeed, flagship cosmological simulations that have been validated at low-redshift, like \eagle\ \citep{schaye_eagle_2015, crain_eagle_2015}, \simba\ \citep{simba}, \illustris\ \citep{vogelsberger_introducing_2014,Vogelsberger2014,Genel2014,Sijacki2015}, and \illustrisTNG\ \citep{Naiman2018,Nelson2018,Marinacci2018,Springel2018,Pillepich2018} fail to pass this threshold, with only a small number of observationally accessible sources at $z>10$. 

One solution is to carry out much larger simulations, but limited to high-redshift. This is a strategy implemented by, e.g., \massiveblack\ \citep{massiveblack}, \bluetides\ \citep{Bluetides-I, Bluetides-II}, and \astrid\ \citep{astrid}. An alternative strategy is to \emph{re-simulate} a range of environments drawn from a very large low-resolution parent simulation \citep[e.g.][]{Crain2009}. This has the advantage of more efficiently allowing us to extend the dynamic range \citep[see discussion in][]{FLARES-I}. These individual re-simulations are also much smaller than single large boxes, allowing wider and more efficient use of HPC systems. The chief disadvantages are the loss of most clustering information and the requirement to carefully understand the weightings of the individual simulations to obtain the correct statistical properties of the galaxy population. Machine learning approaches may allow us to overcome some of these limitations \citep[\textit{e.g.}][]{Lovell_2022, Bernardini22}.

Re-simulations of multiple regions  is utilised in the \flares: First Light And Reionisation Epoch Simulations project. \flares\ combines the validated at $z = 0$ \eagle\ physics model with a re-simulation strategy yielding a much larger effective volume and dynamic range. Compared to the $(100\ {\rm Mpc})^3$ \eagle\ reference simulation, \flares\ contains 10-100$\times$ as many high-redshift galaxies. In this article we use \flares\ to make predictions for the galaxy population at $z>10$, building on earlier work focused at $z=5-10$ \citep{FLARES-I, FLARES-II, FLARES-III, FLARES-IV}. 

This article is organised as follows: in Section \ref{sec:flares} we briefly describe the \flares\ project. In Section \ref{sec:physical} we explore the physical properties of galaxies at $z>10$ and in Section \ref{sec:observational} explore their observational properties, including the UV luminosity function and forecasts for upcoming surveys (\S\ref{sec:observational.uvlf}), the UV continuum slope $\beta$ (\S\ref{sec:observational.beta}), UV emission lines (\S\ref{sec:observational.uvlines}), and the UV - optical colours (\S\ref{sec:observational.uvopt}). Finally, in Section \ref{sec:conc} we present our conclusions.

\section{The First Light And Reionisation Epoch Simulations}\label{sec:flares}

In this study, we make use of the First Light And Reionisation Epoch Simulations \citep[\flares;][]{FLARES-I, FLARES-II}.
\flares\ is a suite of 40 spherical re-simulations, $14\ h^{-1}\, {\rm Mpc}$ in radius, of regions selected from a large $(3.2\ {\rm Gpc})^3$ dark matter only simulation. 
The regions selected to re-simulate span a range of environments: (at $z\approx 4.7$) $\log_{10}(1+\delta_{14}) = [-0.3, 0.3]$\footnote{Where $\delta_{14}$ is the density contrast measured within the re-simulation volume size.} with over-representation of the extremes of the density contrast distribution.

\flares\ adopts the AGNdT9 variant of the \eagle\ simulation project \cite[][]{schaye_eagle_2015, crain_eagle_2015}. The AGNdT9 variant implements a higher heating temperature from active galactic nuclei (AGN) compared to the reference \eagle\ run, thus producing more energetic, less frequent feedback events. We adopt identical resolution to the fiducial \eagle\ simulation, i.e. dark matter and initial gas particle masses of m$_{\mathrm{dm}}=9.7\times10^6\ $M$_{\odot}$ and m$_{\mathrm{g}}=1.8\times10^6$\ M$_{\odot}$, respectively, and a softening length of $2.66 \; \mathrm{ckpc}$. As with the original \eagle\ simulations, we assume $\Omega_{m}$ = 0.307, $\Omega_{\Lambda}$ = 0.693, $h$ = 0.6777 based on results from \citet{planck_collaboration_2014}.

Galaxies in \flares\ are first identified as groups via the Friends-Of-Friends \citep[FOF,][]{davis_evolution_1985} algorithm, and subsequently subdivided into bound groups with the \textsc{Subfind} \citep{springel_populating_2001,dolag_substructures_2009} algorithm. When measuring properties we use 30 kpc radius apertures, centred on the most bound particle of each subgroup.

\subsection{Spectral Energy Distribution Modelling}

Key to making observational predictions is the spectral energy distribution (SED) modelling applied to the simulation outputs. The SED modelling approach is presented in \cite{FLARES-II}, broadly following the approach developed by \cite{Wilkins2013, Wilkins2016a, Bluetides_dust, Wilkins2020}, with modifications to the dust treatment. In short, we begin by associating each star particle with \emph{pure stellar} spectral energy distribution (SED) using v2.2.1 of the Binary Population and Spectral Synthesis \citep[BPASS;][]{BPASS2.2.1} stellar population synthesis model assuming a \cite{chabrier_galactic_2003} initial mass function (IMF). We then associate each star particle with $\mathrm{HII}$ region giving rise to nebular continuum and line emission. Specifically, we follow the approach detailed in \cite{Wilkins2020}, in which the pure stellar spectrum is processed with the \texttt{cloudy} photo-ionisation code \citep{Ferland2017}. We account for the effects of dust attenuation in both the birth clouds of young stellar populations and the interstellar medium (ISM). The latter is accounted for using a line-of-sight (LOS) model similar to that described in \citet{Bluetides_dust}. In this model we treat stellar particles as emitters along a line of sight and account for the attenuation due to gas particles which intersect this LOS. To do this we determine the metal column density and convert this to a dust optical depth using the fitting function for the dust-to-metal ratio presented in \cite{Vijayan2019}. For the attenuation due to the birth cloud component, we scale it with the star particle metallicity, thus assuming a constant dust-to-metal ratio. The proportionality factors for the two components are fixed to match the $z=5$ UV LF from \cite{Bouwens2015a}, $z=5$ UV-continuum slope relations and the [O\textsc{iii}]$\lambda4959,5007$ + H${\beta}$ line luminosity and EW relations at $z=8$ from \cite{deBarros19_OIIIHbeta}. 

\subsection{Comparison to \eagle}

\begin{figure}
	\includegraphics[width=\columnwidth]{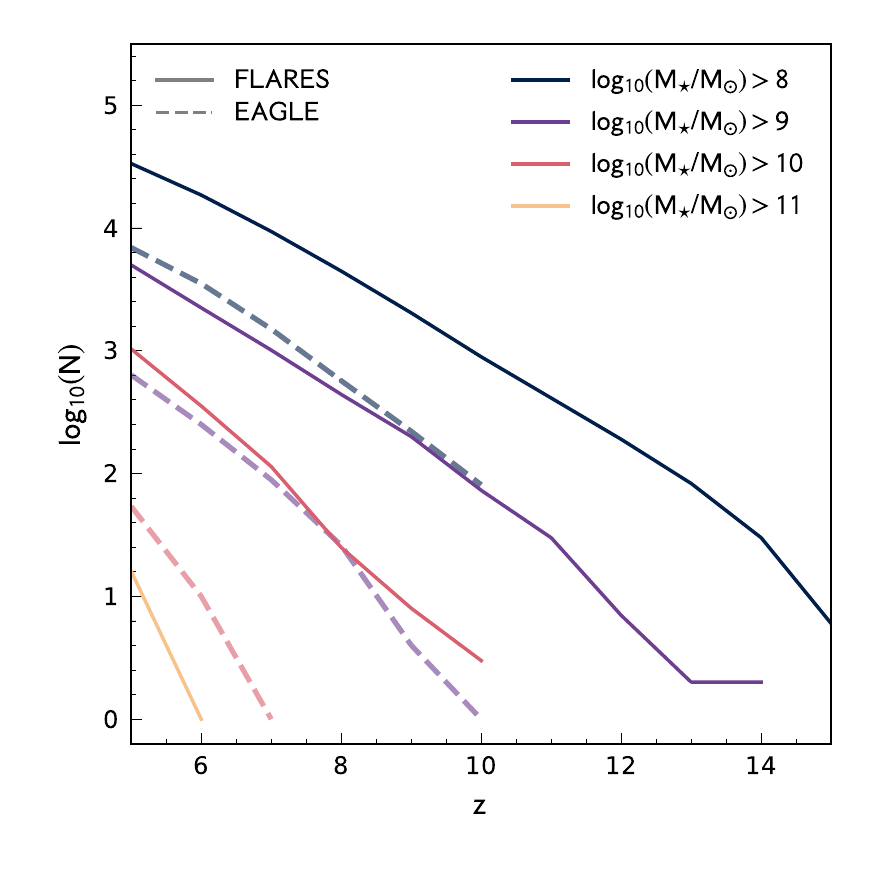}
	\caption{The number of galaxies in \eagle\ (dashed line) and \flares\ (solid line) at $z>5$ above various stellar mass limits.  \label{fig:N}}
\end{figure}

The core objective of \flares\ is to expand both number and dynamic range of simulated galaxies at $z>5$ compared to the \eagle\ reference simulation. The number of galaxies in both \flares\ and \eagle\ with stellar mass greater than $\in\{10^8, 10^9, 10^{10}, 10^{11}\}\ {\rm M_{\odot}}$ at $z>5$ are shown in Fig. \ref{fig:N}. At $z=10$ \flares\ contains 100$\times$ (10$\times$) as many galaxies as \eagle\ with $M_{\star}>10^9\ {\rm M_{\odot}}$ ($M_{\star}>10^8\ {\rm M_{\odot}}$). At $z=10$ \flares\ contains $\sim 1000$ galaxies at $M_{\star}>10^8\ {\rm M_{\odot}}$, dropping to 10 by $z=15$. 

\subsection{Environmental dependence}

\begin{figure}
	\includegraphics[width=\columnwidth]{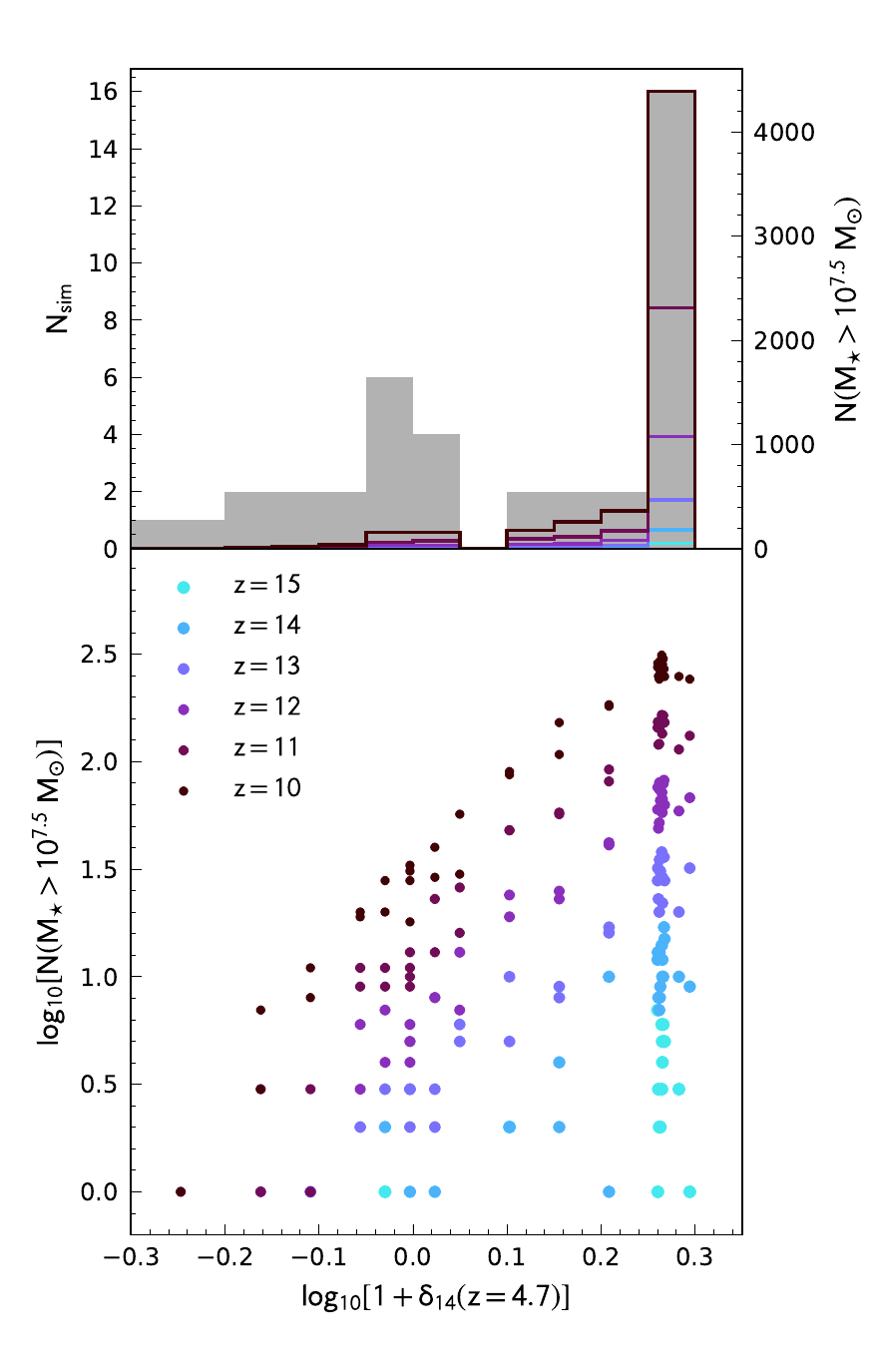}
	\caption{\emph{Bottom-panel:} The number of $M_{\star}>10^{7.5}\ {\rm M_{\odot}}$ galaxies in each individual simulation as a function of density contrast. \emph{Top-panel:} The number of individual simulations (grey filled histogram) and $M_{\star}>10^{7.5}\ {\rm M_{\odot}}$ galaxies in density contrast bins.
	\label{fig:N_delta}}
\end{figure}

A key feature of \flares\ is the explicit simulation of a wide range of environments with $\log_{10}[1+\delta_{14}(z=4.7)] \approx [-0.3, 0.3]$. In \citet{FLARES-I} we showed that the galaxy stellar mass function, and thus the total number of galaxies above a mass threshold, was extremely sensitive to the environment. A consequence of this, and the low number density of galaxies at $z>10$, is that the vast majority of our simulated galaxies are in very over-dense regions as shown in Fig. \ref{fig:N_delta}. 
At $z=15$ only one simulation with $\delta<0.5$ contains any galaxies. 
Since each simulation is appropriately weighted (see \citealt{FLARES-I}) the lack of any galaxies in many density contrast bins should not affect distribution functions like the galaxy stellar mass function or UV luminosity function. However, if galaxy scaling relationships are sensitive to the environment, even the appropriately weighted relations could be biased. Fortunately, \citet{FLARES-I} found no significant evidence of environmental dependence in the key scaling relations.

\section{Physical Properties}\label{sec:physical}

\begin{figure*}
	\includegraphics[width=2\columnwidth]{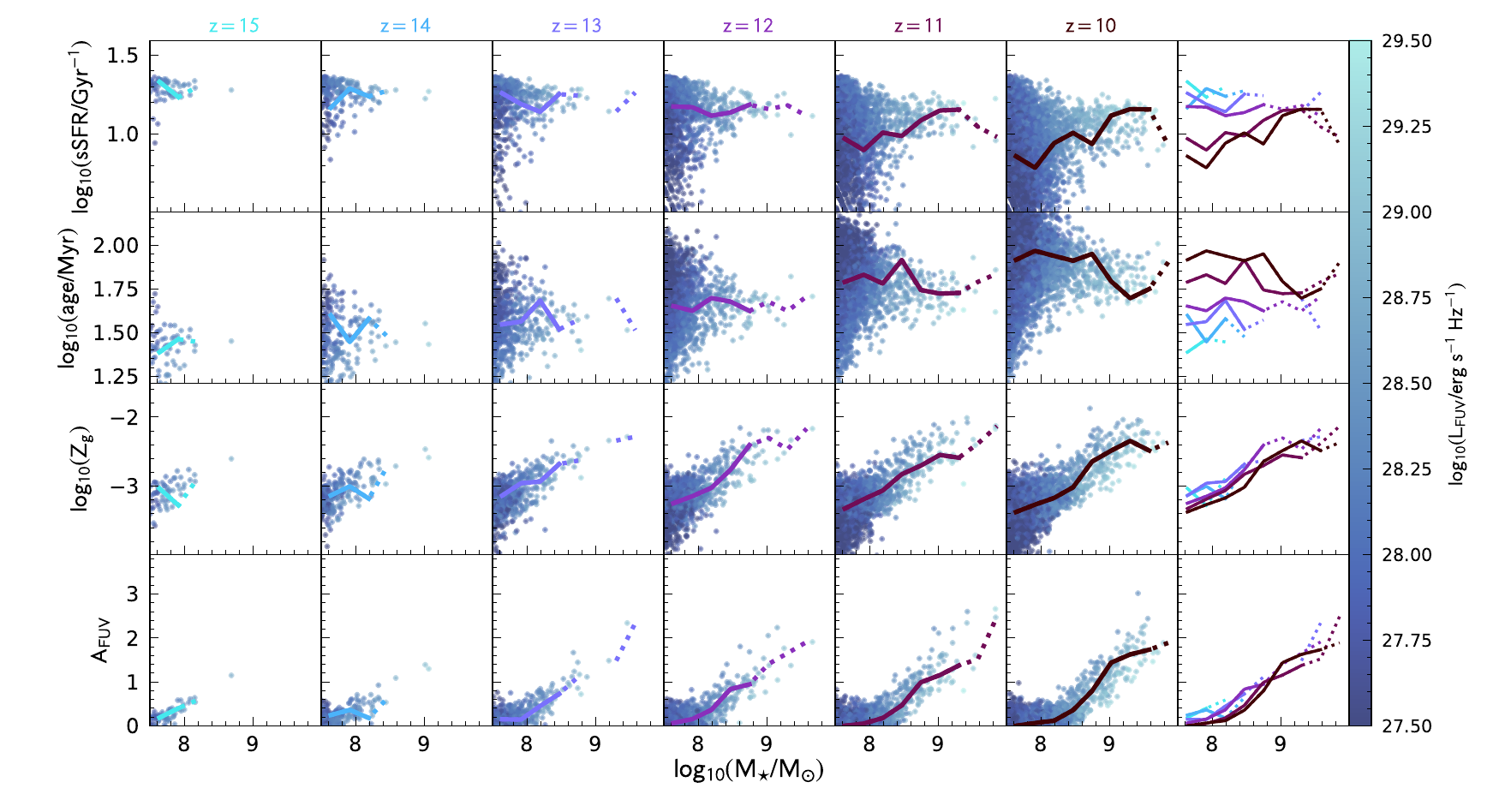}
	\caption{The predicted physical properties as a function of stellar mass at $z=15\to 10$. The line on each panel shows the weighted median, i.e. accounting for the statistical weights of each re-simulation. Where the number of galaxies in each bin falls below 10 the median is denoted by a dotted line. The final panel summarises the redshift evolution of the median. Individual galaxies are coloured by their far-UV luminosity.}
	\label{fig:physical_properties}
\end{figure*}

We begin by exploring a selection of key physical properties of galaxies at $z>10$. In Fig. \ref{fig:physical_properties} we show the relationship between stellar mass and the specific star formation rate, age, gas-phase metallicity, and far-UV dust attenuation. At present, there are no observational constraints for these properties but this should soon change.

The relationship between stellar mass and specific star formation rate is predicted to be fairly flat, though with significant redshift evolution of the normalisation. Similarly, the average age - defined here as the time since its stellar mass was half its current value - is flat with stellar mass but evolves strongly with redshift. At $z=10$, the average age is predicted to be $\approx 20\ {\rm Myr}$ rising to almost $\approx 100\ {\rm Myr}$ at $z=10$. 

On the other hand, the gas phase metallicity $Z_{g}$ shows a significant trend with stellar mass but little redshift evolution. Similarly we see a strong trend between the far-UV attenuation and stellar mass but little redshift evolution. Given the gas phase metallicity trends this is unsurprising since the attenuation in \flares\ is related to surface density of metals. 

\subsection{Mass-to-light ratio}\label{sec:mtol}

\begin{figure*}
	\includegraphics[width=2\columnwidth]{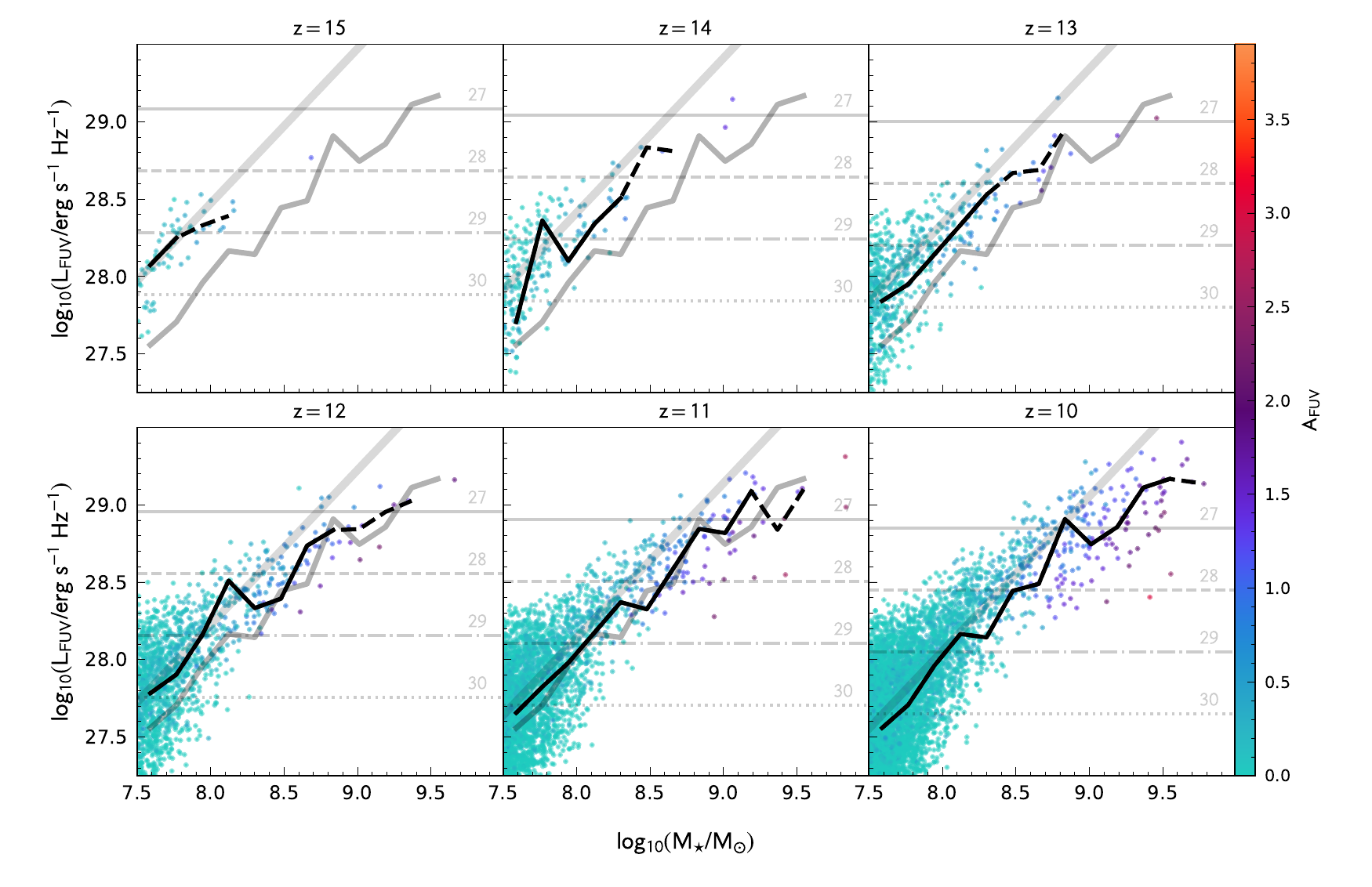}
	\caption{The relationship between stellar mass and the dust-attenuated (observed) far-UV luminosity at $z=15\to 10$. Points are colour-coded by the far-UV attenuation $A_{\rm FUV}$. The black line denotes the weighted median at each redshift with the line becoming dashed when the number of galaxies in each bin falls below 10. The thin grey line shows the median at $z=10$. The thick grey line shows a weighted linear fit. Horizontal lines denote the corresponding apparent magnitude.\label{fig:L}}
\end{figure*}

One of the most fundamental properties is the mapping between stellar mass and the (dust-attenuated) far-UV luminosity. This encodes the star formation and metal enrichment history of each galaxy in addition to reprocessing by dust and gas. This relationship is shown in Fig. \ref{fig:L}. At $M_{\star}<10^{8.5}\ {\rm M_{\odot}}$ this relationship is close to linear; at high-masses, however, the luminosity falls below the linear expectation. This is predominantly due to the effects of dust, but is also affected by the higher stellar metallicities in the most massive galaxies. The scatter in this relationship is $\approx 0.2\ {\rm dex}$ independent of redshift and stellar mass. 

\section{Observational properties}\label{sec:observational}

\begin{figure}
 	\includegraphics[width=\columnwidth]{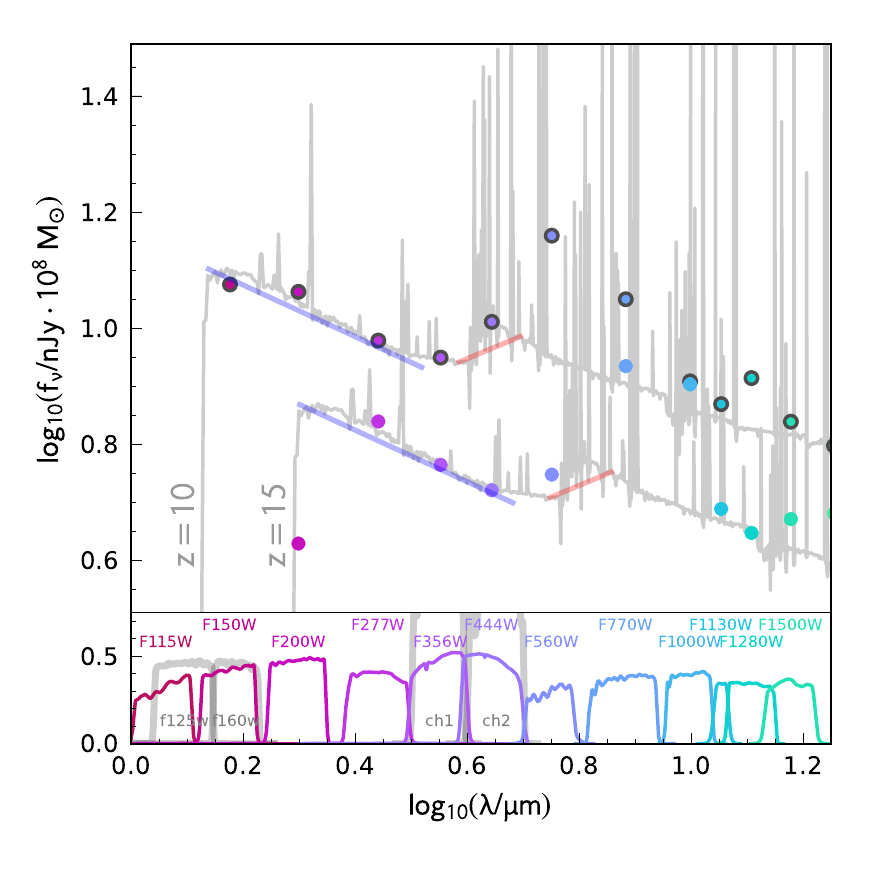}
	\caption{The observed spectral energy distribution of a star forming galaxy at $z=10$ and $z=15$ alongside key \hubble, \spitzer, \jwst/NIRCam, and \jwst/MIRI filter transmission functions. Coloured points denote the predicted fluxes in each of the NIRCam and MIRI bands highlighting the impact of nebular emission in the rest-frame optical. The blue and red lines denote the UV continuum slope and Balmer break respectively.
	\label{fig:sed}}
\end{figure}

We now turn our attention to some of the properties of galaxies that can be observed at $z>10$. \jwst\ will, for the first time, provide deep >2 $\mu$m imaging and spectroscopy allowing us to measure several properties including the rest-frame UV luminosity (and thus luminosity function), the UV continuum slope, UV emission lines, and potentially even UV - optical colours via MIRI imaging. Model spectral energy distributions of star forming galaxies at $z=10$ and $z=15$ are shown, alongside the \jwst\ filter transmission functions in Fig. \ref{fig:sed}.

\subsection{Far-UV luminosity function}\label{sec:observational.uvlf}

The rest-frame far-UV luminosity function (LF) is one of the key statistical descriptions of the galaxy population at high-redshift. This is predominantly due to its accessibility but also the fact that the observed UV light traces both unobscured star formation \citep[][]{KE2012, Wilkins2019} and the production of ionising photons \citep[][]{Wilkins2016}. The far-UV LF has now been measured extensively to $z\sim 8$ with tentative constraints at $\sim 10$ \citep[e.g][]{McLeod2016, Oesch2018, Finkelstein2022} and more recently at $z\sim 13$ \citep{Harikane22}. 

\begin{figure*}
	\includegraphics[width=2\columnwidth]{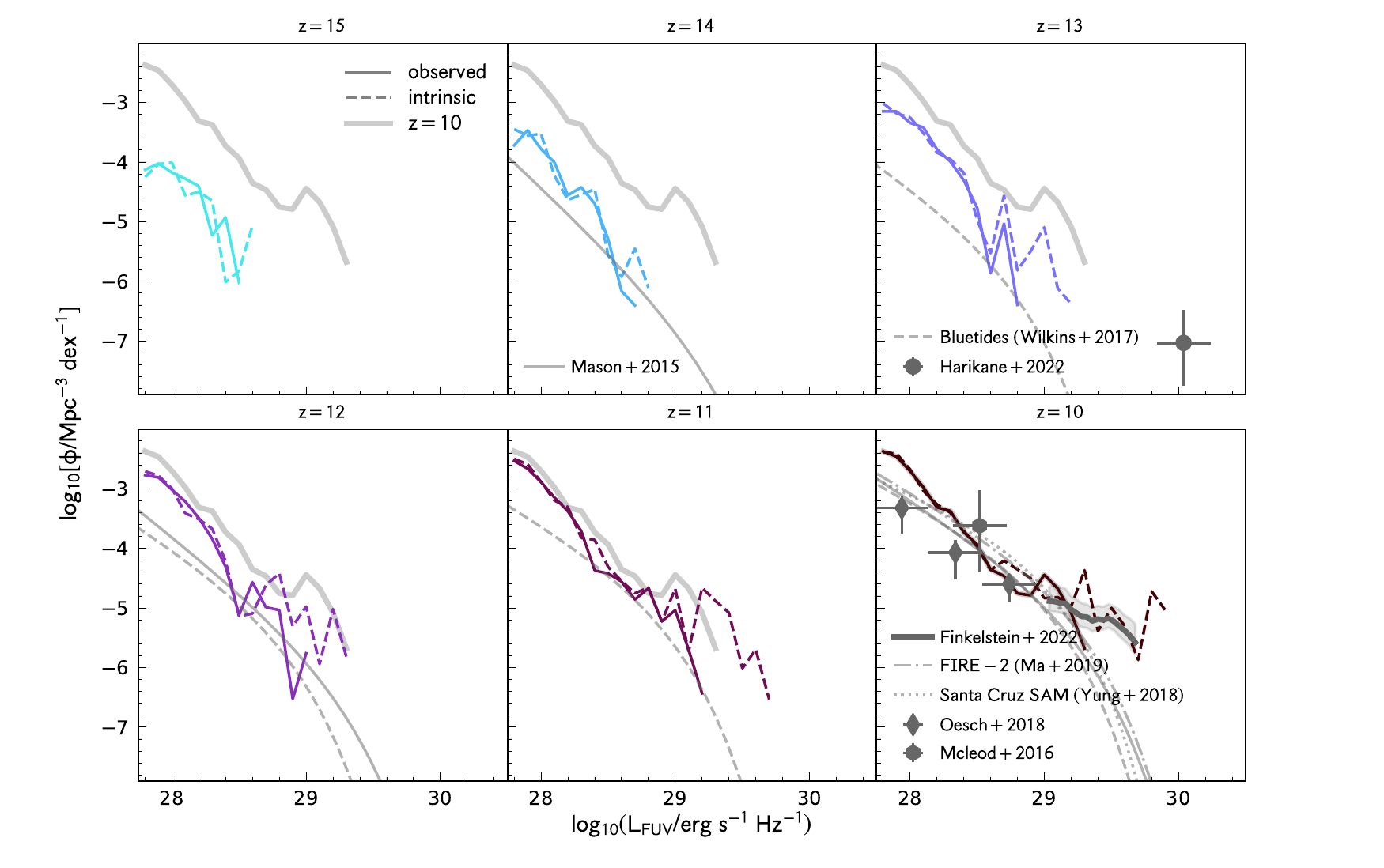}
	\caption{The observed (dust-attenuated, sold line) and intrinsic (dashed line) far-UV luminosity function at $z=15\to 10$ predicted by \flares. The thick grey line denotes the $z=10$ prediction to more clearly demonstrate the redshift evolution. Observational constraints at $z=10$ and $z=13$ from \citet{McLeod2016}, \citet{Oesch2018}, \citet{Finkelstein2022}, and \citet{Harikane22} are also shown. Model predictions from \citet{Mason15}, \citet{Bluetides-II}, \citet{YungI}, and \citet{Ma2019} are also shown where available. These \flares\ binned luminosity functions are available in Table \ref{tab:uvlf}. \label{fig:uvlf}}
\end{figure*}

In Fig. \ref{fig:uvlf} we show the far-UV luminosity predicted by \flares\ at $z=15\to 10$  alongside both these observational constraints and other model predictions. We show both the \emph{observed} (dust-attenuated) and \emph{intrinsic} distribution functions. The far-UV LF follows the familiar steep decline with luminosity seen at low-redshift, dropping by roughly a factor of $10^3$ from $L=10^{28}\to 10^{29}\ {\rm erg\ s^{-1}\ Hz^{-1}}$. The number density of sources also evolves strongly, increasing by $\sim 10\times$ from $z=14\to 10$ with stronger evolution at the bright end. At low-luminosity ($L<10^{28.5}\ {\rm erg\ s^{-1}\ Hz^{-1}}$) the impact of dust is small leaving the intrinsic and observed LF similar. However, as noted previously brighter, more massive galaxies, increasingly have stronger dust attenuation leading to a divergence in the predictions. 

The $z=10$ panel shows current observational constraints from \citet{McLeod2016}, \citet{Oesch2018}, and \citet{Finkelstein2022} all based on \hubble/WFC3 observations. While the observational uncertainties are large \citet{McLeod2016} and \citet{Oesch2018} bracket the \flares\ predictions; the \citet{Oesch2018} constraints falling slightly below the predictions at low luminosities. The \citet{Finkelstein2022} constraints lie at the bright end of our predictions where dust attenuation is predicted to become important. These constraints agree better with the \flares\ intrinsic predictions, possibly suggesting that too much dust is assumed in \flares\ at this redshift. 

Shown on the $z=13$ panel are the recent observational constraints from \citet{Harikane22}. This study used observations from  Hyper Suprime-Cam, VISTA, and \emph{Spitzer} of the COSMOS and SXDS fields to identify a pair of bright sources with spectral energy distribution consistent with $z\sim 13$ star forming galaxies. In addition, one source has a tentative line detection consistent with [O\textsc{iii}]88$\mu$m  at $z=13.27$, in-line with its photometric redshift. If real these sources suggest little evolution in the bright end of the UV luminosity function from $z\sim 10\to 13$. \flares\ contains objects with a similar space density as these sources but with dust -attenuated luminosities around a factor of $10\times$ smaller. As noted, at $z>10$ the \flares\ dust model is likely to become increasingly unreliable since it is based on modelling calibrated at lower-redshift. However, even using intrinsic luminosities, sources with a similar space density in \flares\ are still around $5\times$ fainter, suggesting significant remaining tension. 

Fig. \ref{fig:uvlf} also shows a comparison with other model predictions at $z\ge 10$ including the semi-empirical model of \citet{Mason15}, the Santa Cruz semi-analytical model \citep{YungI}, the large volume cosmological hydrodynamical simulation \bluetides\ \citep{Bluetides-II}, and the high-resolution \textsc{fire-2} simulations \citep{Ma2019}. While there is good agreement at $z=10$, the agreement with \citet{Mason15} and \bluetides\ breaks down at high-redshift; \flares\ predicts a similar density of bright galaxies but consistently predicts more faint galaxies. 


\begin{table}

\begin{tabular}{ccccccc}
\hline
$\log_{10}(L_{\rm FUV})$ & \multicolumn{6}{c}{$\phi/{\rm Mpc^{-3}\ dex^{-1}}$} \\
${\rm erg\ s^{-1}\ Hz^{-1}}$ & $z=15$ & $z=14$ & $z=13$ & $z=12$ & $z=11$ & $z=10$ \\
\hline\hline
27.8 & -4.16 & -3.59 & -3.14 & -2.78 & -2.53 & -2.37 \\
28.0 & -4.10 & -3.64 & -3.35 & -3.02 & -2.92 & -2.67 \\
28.2 & -4.67 & -4.47 & -3.64 & -3.45 & -3.29 & -3.19 \\
28.4 & -5.07 & -4.71 & -4.33 & -4.29 & -4.07 & -3.65 \\
28.6 & -7.04 & -6.01 & -4.95 & -4.83 & -4.60 & -4.33 \\
28.8 & -7.45 & -6.71 & -6.29 & -5.02 & -4.88 & -4.56 \\
29.0 & - & -7.45 & -6.12 & -6.06 & -5.14 & -4.64 \\
29.2 & - & -6.20 & -7.04 & -6.12 & -6.04 & -4.82 \\
29.4 & - & - & - & - & -7.22 & -6.68 \\
\hline
\end{tabular}
\caption{The space density of galaxies at $z=10\to 15$ predicted by \flares. \label{tab:uvlf} }
\end{table}

\subsubsection{JWST Cycle 1 forecasts}

As noted in the introduction the $z>10$ galaxy population will soon be accessible via several deep imaging surveys conducted by \jwst. Using our luminosity function predictions we can now forecast the number of sources expected for each survey and the eventual constraints on the $z=10-15$ LF.

\begin{figure}
	\includegraphics[width=\columnwidth]{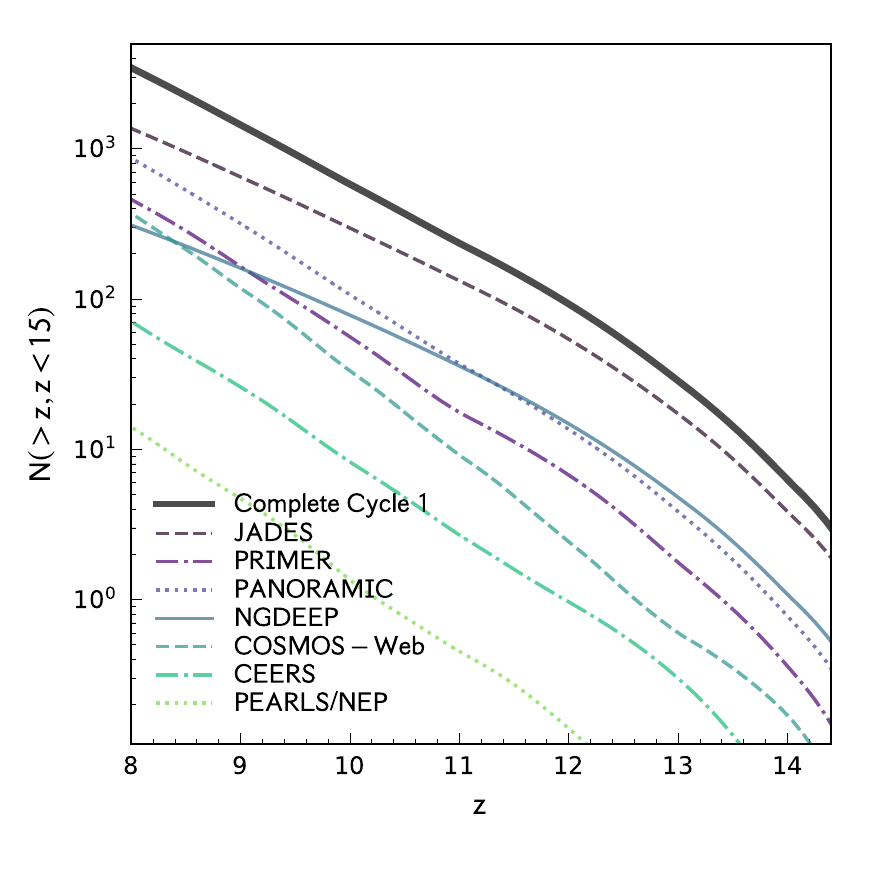}
	\caption{The cumulative number of galaxies $N(>z, z<15)$ predicted by FLARES for various JWST Cycle 1 GO, ERS, and GTO surveys assuming 100\% completeness down to the $10\sigma$ point-source F277W depth. The thick black line denotes the prediction for all surveys combined. \label{fig:survey_forecast}}
\end{figure}

We begin, in Fig. \ref{fig:survey_forecast}, by presenting forecasts for the cumulative number of sources accessible to various JWST Cycle 1 GO, ERS, and GTO programmes. These  include: CEERS, NG-DEEP\footnote{Formerly Webb DEEP.}, PRIMER, COSMOS-Web\footnote{Formerly COSMOS-Webb.}, JADES, the Northern Ecliptic Pole element of PEARLS\footnote{Formerly the Webb Medium Deep Fields programme (PI Windhorst).}, and PANORAMIC. These predictions assume 100\% completeness down to the $10\sigma$ point-source F277W depth, which in reality is likely to be difficult to achieve. The approximate depths/areas for each survey were provided by each programme PI except for JADES which are taken from \citet{Williams2018}. Across all 7 programmes we predict $\sim 500$, 85, and 6 galaxies at $z>10$, $z>12$, and $z>14$ respectively. JADES is predicted to dominate these numbers contributing around half of expected sources at these redshifts.

\begin{figure*}
	\includegraphics[width=2\columnwidth]{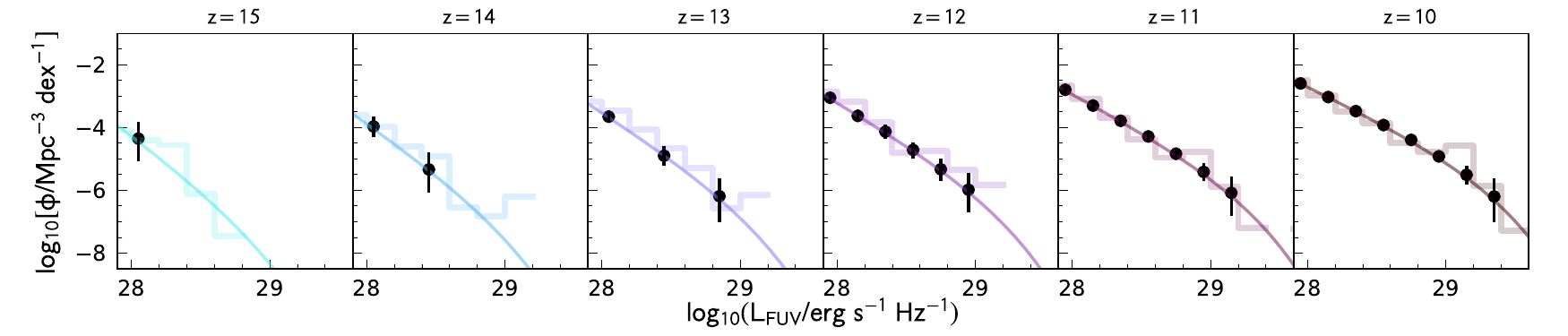}
	\caption{Predicted constraints on the UV $z=15\to 10$ UV luminosity function from combined JWST Cycle 1 GO, ERS, and GTO surveys. Points show the forecast binned LF constraints. The stepped line shows the \flares\ binned LF constraints while the smooth lines show Schechter function fits to the \flares\ LF. \label{fig:lf_forecast}}
\end{figure*}

In Fig. \ref{fig:lf_forecast} we then make forecasts for the $z=15\to 10$ luminosity function for the combination of the Cycle 1 programmes. The result is strong constraints at $z=10$, comparable to the current $z=7$ constraints from \hubble's entire campaign \citep{BouwensUVLF2021}. While these constraints progressively weaken toward higher redshift subsequent observations throughout \jwst's tenure should ultimately enable useful constraints to $z\sim 15$ and potentially beyond. Crucially this will allow us to differentiate between the different model predictions in this era.

\subsection{Sizes}\label{sec:observational.sizes}

\begin{figure}
	\includegraphics[width=\columnwidth]{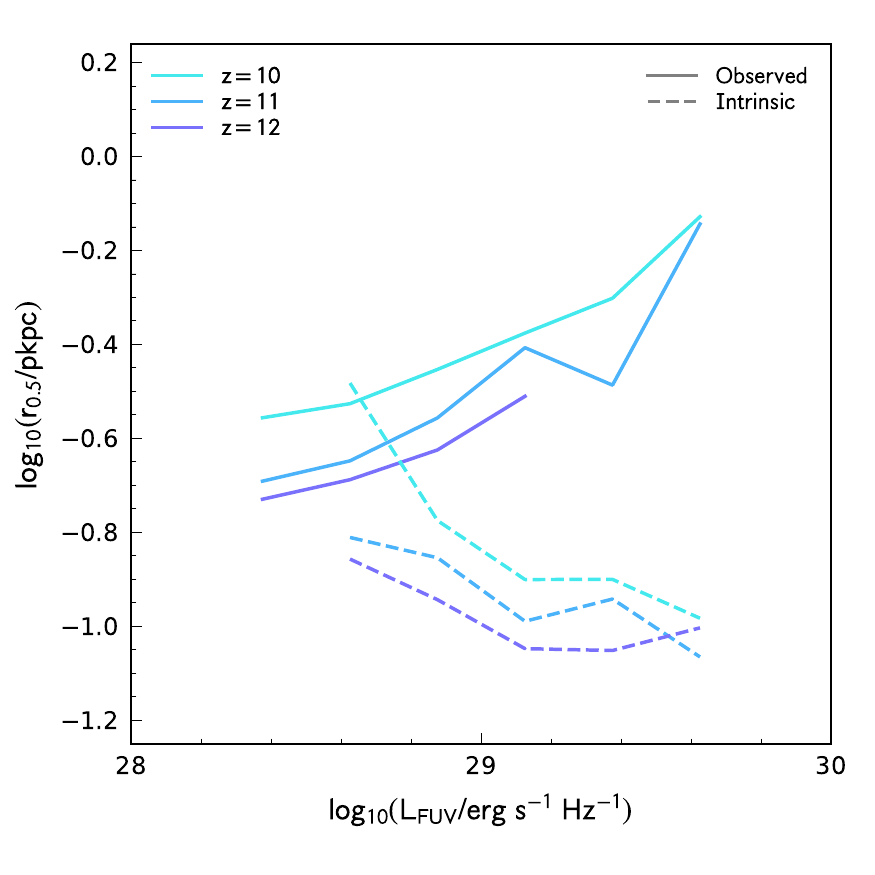}
	\caption{The redshift evolution of the size - luminosity relation predicted by \flares\ in the redshift range $z=10-12$. The lines represent the weighted 50th percentile of the galaxy distribution weighted using the \flares\ weighting scheme. Solid lines show the observed size including the effects of dust attenuation, while dashed lines show the intrinsic stellar emission. Observed sizes are measured using a non-parametric pixel approach while intrinsic sizes use a particle based approach. \label{fig:size}}
\end{figure}

In Fig. \ref{fig:size} we present measurements of galaxy half light radius in the far-UV for $z=10-12$. The intrinsic size measurements are derived from the particle distribution, while the observed size measurements use a non-parametric pixel approach in which size is derived from the non-contiguous pixel area containing half the galaxy's total luminosity. The latter approach well encompasses the clumpy natures of high redshift galaxies \citep[e.g][]{Jiang_2013, Bowler2017}. The redshift range is limited by the number of galaxies in \flares\ with a sufficient number of stellar particles to make robust size measurements ($N_\star\geq100$). As in \cite{FLARES-IV} we impose a 95 percent completeness limit in each redshift bin. 

Intrinsically the high redshift galaxy population is extremely compact with a negative size-luminosity relation. However, the bright central regions of these compact galaxies are also efficiently seeding their surroundings with metals, even at this early epoch. This seeding leads to strong dust obscuration in the brightest regions, and thus a large increase in the observed size and a positive observed size-luminosity relation. The normalisation of the observed size-luminosity relation quickly evolves at these high redshifts, with an increase of $\sim0.1$ dex from $z=12$ to $z=10$ driven by increasing obscuration of the brightest regions due to the formation of dust in the highly star forming cores of these bright galaxies. \jwst\ will not only be able to probe this obscured size-luminosity relation in the rest frame far-UV, the reddest NIRCam filters will also be able to probe deeper into the increasingly unobscured size-luminosity relation at longer wavelength. This will provide a valuable view into the intrinsic size-luminosity relation and its negative slope. 

\subsection{The UV continuum slope}\label{sec:observational.beta}

The most accessible spectral diagnostic available at high-redshift is the UV continuum slope $\beta$: as can be seen in Fig. \ref{fig:sed} the rest-frame UV continuum to $\lambda = 350$ nm is accessible to NIRCam to $z\approx 15$ and can be measured with 3-4 of NIRCam's wide filters.  While primarily an indicator of dust attenuation the UV continuum slope is also sensitive to the star formation and metal enrichment history, the Lyman continuum escape fraction, the initial mass function, and our understanding of stellar evolution and atmospheres\footnote{In a theoretical context this is encapsulated in stellar population synthesis models.} \citep[e.g][]{Wilkins2013, MassiveBlack_UVslopes}. 

\begin{figure*}
	\includegraphics[width=2\columnwidth]{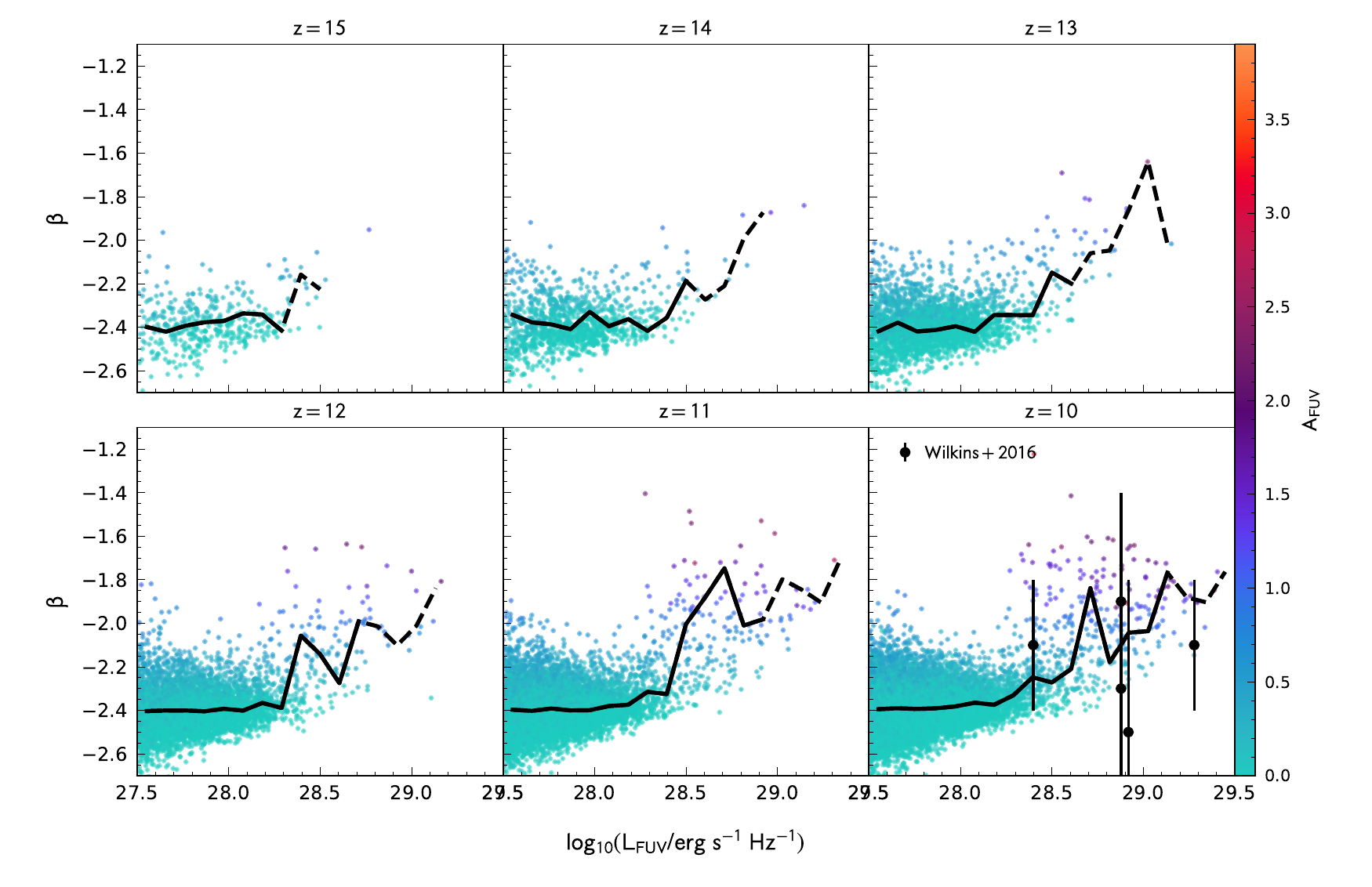}
	\caption{The rest-frame UV continuum slope $\beta$ as a function of far-UV luminosity at $z=15\to 10$. Individual galaxies are coloured coded by their far-UV attenuation. The solid (dashed) black line shows the median value for bins with $\ge 10$ ($<10$) galaxies. Observational constraints from \citet{Wilkins_z10UVslopes} are also shown at $z=10$. \label{fig:beta}}
\end{figure*}

Fig. \ref{fig:beta} shows predictions for $\beta$ from $z=15\to 10$ colour coded by the rest-frame far-UV attenuation. At $L_{\rm FUV}< 10^{28.5}\ {\rm erg\ s^{-1}\ Hz^{-1}}$ ($M_{\rm FUV}>-19.5$) the slope is $\approx -2.4$ with little evolution with redshift. At $L_{\rm FUV}> 10^{28.5}\ {\rm erg\ s^{-1}\ Hz^{-1}}$ the slope progressively reddens due to the increasing dust attenuation. At $L_{\rm FUV}> 10^{29}\ {\rm erg\ s^{-1}\ Hz^{-1}}$ the average slope has reddened to $\approx -2$. Fig. \ref{fig:beta} also shows observational constraints from \citet{Wilkins_z10UVslopes}; while the uncertainties are large these observations are consistent with our predictions.

\begin{figure}
	\includegraphics[width=\columnwidth]{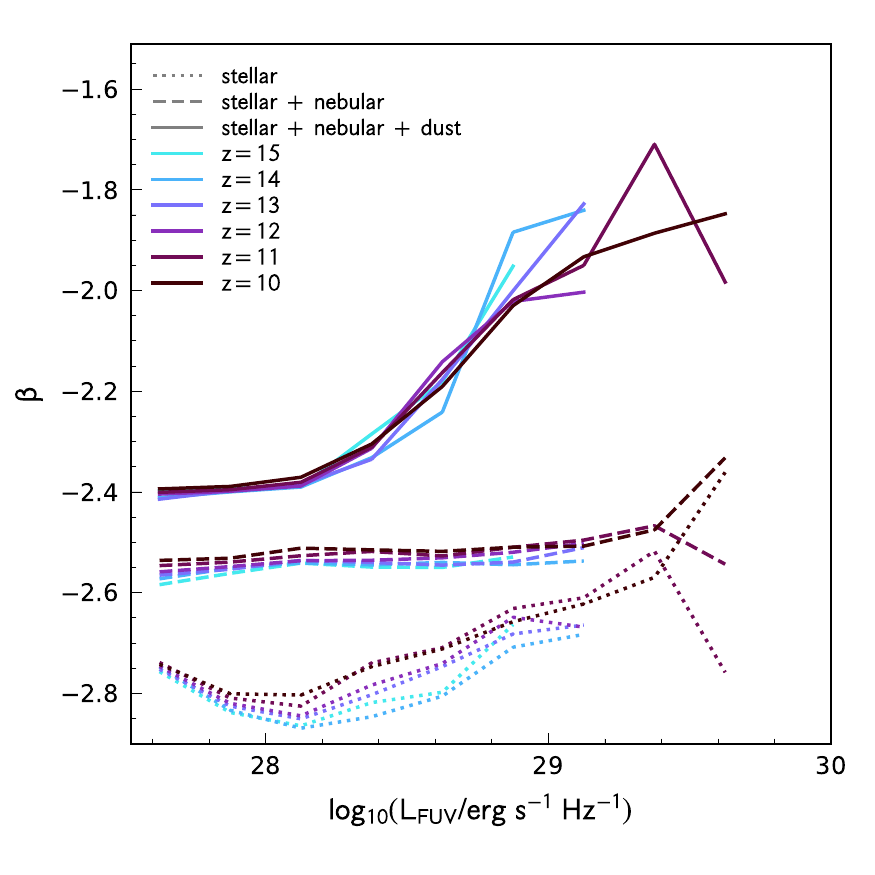}
	\caption{The average UV continuum slope as a function of UV luminosity and redshift. The dotted line shows the UV continuum slope predicted from just the pure stellar emission, the dashed line including nebular emission, and the solid line also including dust (i.e. that which would be observed). \label{fig:beta2}}
\end{figure}

The origin of the UV continuum slope is explored in more detail in Fig. \ref{fig:beta2}. Here we show the median $\beta$ for unprocessed starlight (dotted line), starlight with reprocessing by gas (dashed line), and reprocessing with gas and dust (solid line, the same as that shown in Fig. \ref{fig:beta}). With no reprocessing the predicted slopes are $\approx -2.8$ at low-luminosity rising to $\approx -2.5$ at  $L_{\rm FUV}\approx 10^{29.5}\ {\rm erg\ s^{-1}\ Hz^{-1}}$ with some weak ($\Delta\beta=0.1$) evolution with redshift $z=15\to 10$. These trends reflect variation in the star formation and metal enrichment history with the brightest galaxies typically having higher metallicities. The addition of nebular (continuum) emission reddens $\beta$. As the impact of nebular emission is strongest at low metallicity this has the effect of flattening the previous trend with luminosity and redshift leaving galaxies with $\beta \approx 2.5$. The addition of dust has an impact at all luminosities though this is most pronounced at $L_{\rm FUV}> 10^{29}\ {\rm erg\ s^{-1}\ Hz^{-1}}$ where dust is predicted to redden the slope by $\approx 0.5$. 

As noted previously the impact of dust attenuation in \flares\ is particularly uncertain at $z>10$. Recent luminosity function constraints \citep[i.e.][]{Finkelstein2022, Harikane22} tentatively suggest the presence of too much dust attenuation in \flares. Precise constraints on the UV continuum slope from \jwst\ - and ideally ALMA dust continuum observations -  in bright ($L_{\rm FUV}> 10^{29}\ {\rm erg\ s^{-1}\ Hz^{-1}}$) should allow us to determine whether this is the case.  

\subsection{UV emission lines}\label{sec:observational.uvlines}

In addition to deep near-IR imaging, \jwst\ will also be able to obtain deep near-IR spectroscopy utilising NIRSpec, NIRCam, and NIRISS\footnote{MIRI also has spectroscopic capabilities but has much lower sensitivity and only target galaxies individually.}. NIRSpec provides both a multi-object and integral field unit spectroscopy at 0.6-5.3$\mu$m, while NIRCam and NIRISS together provide wide field slit-less spectroscopy across the near-IR. At $z<10$ this enables the observation of various strong optical lines. However, at $z>10$ most of the strong lines fall outside the accessible range leaving a handful of weaker UV lines. Most prominent amongst these is the [C\textsc{iii}],C\textsc{iii}]$\lambda\lambda 1907,1909$\AA\ doublet for which a handful of detections are already available at $z>6$ \citep{Stark2015, Stark2017, Topping2021}. 

\begin{figure}
	\includegraphics[width=\columnwidth]{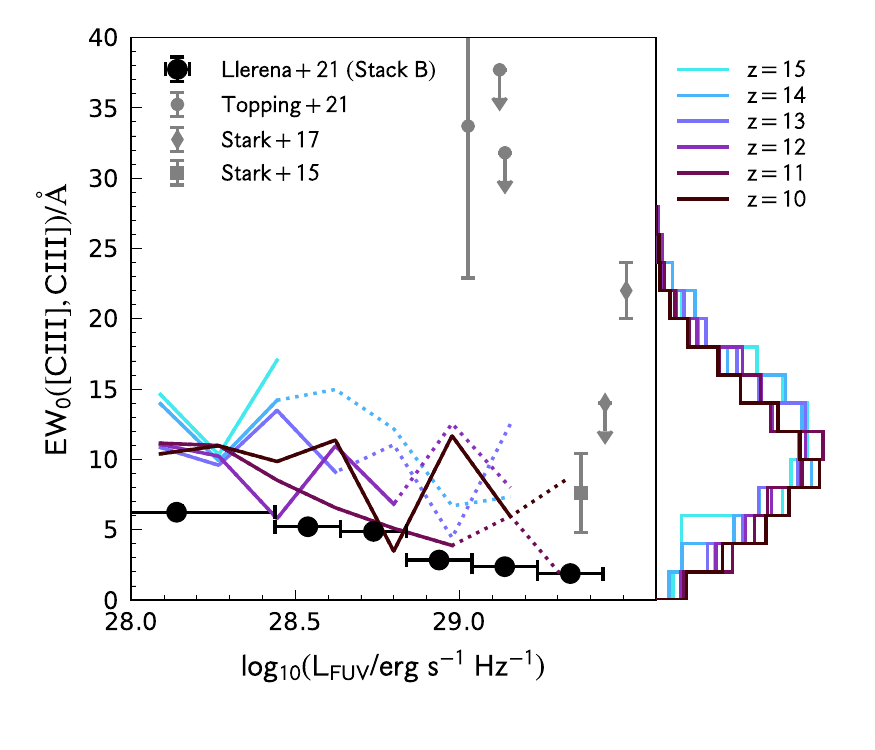}
	\caption{The distribution of [C\textsc{iii}],C\textsc{iii}]$\lambda\lambda 1907,1909$\AA\ equivalent widths at $z=10-15$ predicted by \flares. Also shown are stacked results at $z\approx 3$ from the VANDELS survey \citep{Llerena2022} and individual objects at $z>6$ \citep{Stark2015, Stark2017, Topping2021}. \label{fig:CIII}}
\end{figure}

Predictions from \flares\ for the rest-frame equivalent width distribution of [C\textsc{iii}],C\textsc{iii}] are presented in Fig. \ref{fig:CIII} as a function of the observed (dust attenuated) far-UV luminosity. Equivalent widths show a weak decline to higher luminosity and lower-redshift. The median value is $\approx 10$\AA\ with the tail of the distribution reaching beyond 20\AA. Unsurprisingly, given the younger age, these predictions are offset to higher equivalent widths than observations at lower redshift \citep[e.g][]{Maseda17, Llerena2022}. Constraints at $z>6$ include a handful of detections and upper-limits and are likely biased due to their selection method. However, two of the detections: EGS-zs8-1 \citep{Stark2017} and AEGIS-33376 \citep{Topping2021} have equivalent widths at the upper extreme of the predicted distribution suggesting some possible tension. This may reflect some of the simplifying assumptions used in our modelling. For example, \citet{Wilkins2020} showed that the equivalent width of [C\textsc{iii}],C\textsc{iii}] is strongly sensitive to the ionisation parameter and hydrogen density, for which we adopted single fiducial values. 

\subsection{UV - Optical colours}\label{sec:observational.uvopt}

The near-UV - optical colour is another key spectral diagnostic of galaxies, its measurement providing insights into the star formation and metal enrichment history, dust attenuation, and nebular emission of galaxies. In the context of $z>6$ galaxies the near-UV - optical colour is chiefly impacted by nebular line emission and can be used to infer the strength of combination of the H$\beta$ and [O\textsc{iii}] lines \citep[e.g][]{DeBarros2019, Endsley2021}. Where a spectroscopic redshift is available it is sometimes possible to avoid strong line emission providing "clean" constraints on the strength of the Balmer break \citep[e.g][]{Hashimoto2018} and thus a more accurate constraint on the star formation and metal enrichment history. For a wider introduction to the break in the context of the high-redshift Universe see Wilkins et al. \emph{in-prep}.

\begin{figure}
	\includegraphics[width=\columnwidth]{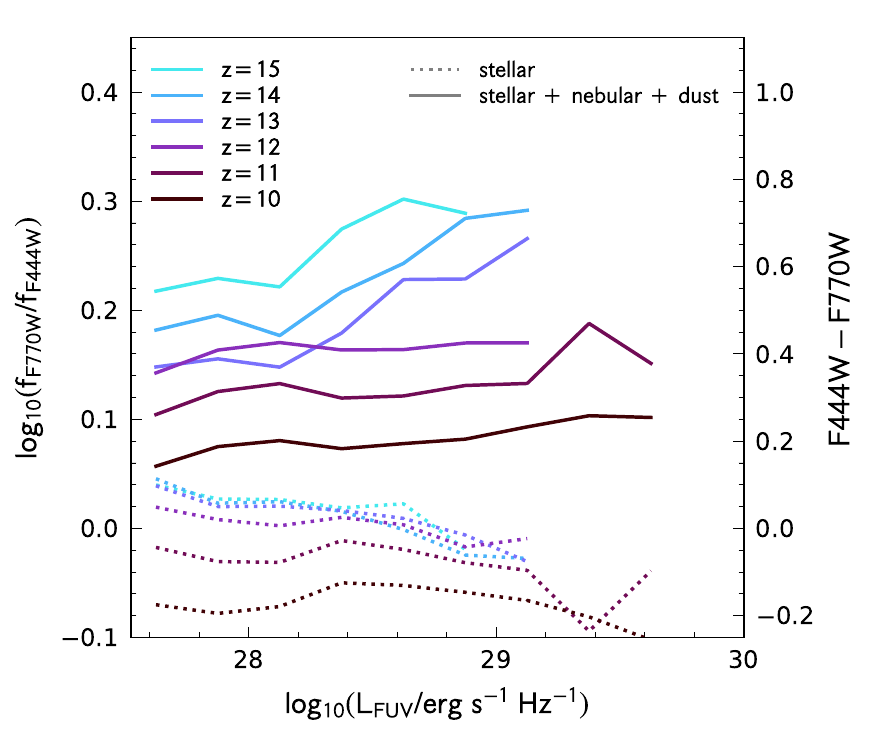}
	\caption{The average NIRCam/F444W - MIRI/F770W colour as a function of UV luminosity for both the pure stellar emission (dotted line) and including nebular emission and dust (solid line). \label{fig:miri}}
\end{figure}

In principle \jwst\ can observe the rest-frame optical to $z=15$ and beyond. However, at $z>11$ the optical falls beyond the range accessible to \jwst's near-IR instruments and thus requires MIRI observations. Several blind-field cycle 1 surveys will simultaneously collect both multi-band NIRCam and MIRI imaging. For MIRI F770W is the most popular choice and with this in mind we present the predicted F444W-F770W colour in Fig. \ref{fig:miri}. We do this for  both the pure stellar photometry and the photometry including nebular emission and dust. In both cases there is little trend with the UV luminosity. The addition of nebular emission (and dust) has the impact of significantly reddening the colour by $\approx 0.4$ mag at $z=10$ increasing to $\approx 0.7$ mag at $z=15$. This reflects both the increasing optical line equivalent widths but also the lines that fall within the F770W band.

While in principle, MIRI observations, when combined with NIRCam, should thus allow us to constrain optical line emission at $z>10$, unfortunately MIRI has both a lower sensitivity\footnote{In 10 ks NIRCam can reach in 9.1 nJy and 23.6 nJy in the F200W and F444W bands respectively while for the MIRI F560W and F770W bands the flux sensitivity is 130 nJy and 240 nJy.} and smaller field-of-view than NIRCam resulting in a much reduced survey efficiency. Blind field cycle 1 programmes (i.e. PRIMER, COSMOS-Web, CEERS) obtaining both NIRCam and MIRI F770W observations typically reach depths F770W 3 magnitudes shallower than the deepest NIRCam observations and only over less than half the total area. Combined with the expected surface densities and flux distributions it is then unlikely, even with stacking, that MIRI will yield useful constraints at $z>10$. However, there is the possibility of obtaining deep MIRI imaging of individual bright targets in later cycles.

\section{Conclusions}\label{sec:conc}

In this article we have presented theoretical predictions for the properties of galaxies at $z=10-15$ from the \flares: First Light And Reionisation Epoch Simulations. These are amongst the first predictions from a cosmological hydrodynamical simulation calibrated at $z = 0$ at these redshifts enabled by the unique simulation strategy adopted by \flares. Our major findings are:
\begin{itemize}

    \item Specific star formation rates and ages show little trend with stellar mass though evolve strongly with redshift. However, gas-phase metallicities and dust attenuation rapidly increase with stellar mass but show little redshift evolution. 
    \item The far-UV luminosity function continues its evolution from lower-redshift with the luminosity density predicted to drop by $\sim 10\times$ from $z=10\to 14$. At $z=10$ the predictions are consistent with observational constraints from \citet{McLeod2016}, \citet{Oesch2018}, and \citet{Finkelstein2022} though favour less dust attenuation. \flares\ contains galaxies with a similar space density to those recently identified by \citet{Harikane22} but $\sim 5-10\times$ fainter depending on whether dust attenuation is included. Similarly agreement with other models is good at $z=10$ but diverges to higher-redshift with \flares\ predicting more faint galaxies than other models. Based on these predictions in cycle 1 alone \jwst\ should identify $\sim$600, 100, and 6 galaxies at $z>10$, $12$, and $14$ respectively providing robust constraints on the LF to $z\sim 13$.
    \item \flares\ predicts little redshift evolution in the relationship between the UV continuum slope $\beta$ and UV luminosity. The brightest galaxies are predicted to be moderately reddened ($\Delta\beta\approx 0.5$) by dust.
    \item UV-optical colours probed by NIRCam and MIRI are likely to be dominated by nebular emission though will be hard to measure due to MIRI's much lower sensitivity. 
\end{itemize}

\section*{Acknowledgements}

We dedicate this article to healthcare and other essential workers,  the teams involved in developing the vaccines, and to all the parents who found themselves having to home-school children while holding down full-time jobs. We thank the \eagle\, team for their efforts in developing the \eagle\, simulation code.  This work used the DiRAC@Durham facility managed by the Institute for Computational Cosmology on behalf of the STFC DiRAC HPC Facility (\url{www.dirac.ac.uk}). The equipment was funded by BEIS capital funding via STFC capital grants ST/K00042X/1, ST/P002293/1, ST/R002371/1 and ST/S002502/1, Durham University and STFC operations grant ST/R000832/1. DiRAC is part of the National e-Infrastructure. CCL acknowledges support from the Royal Society under grant RGF/EA/181016. DI acknowledges support by the European Research Council via ERC Consolidator Grant KETJU (no. 818930). The Cosmic Dawn Center (DAWN) is funded by the Danish National Research Foundation under grant No. 140. We also wish to acknowledge the following open source software packages used in the analysis: \textsc{Numpy} \citep{Harris2020_numpy}, \textsc{Scipy} \citep{2020SciPy-NMeth}, and \textsc{Matplotlib} \citep{Hunter:2007}. This research made use of \textsc{Astropy} \url{http://www.astropy.org} a community-developed core Python package for Astronomy \citep{astropy:2013, astropy:2018}. Parts of the results in this work make use of the colormaps in the \textsc{CMasher} package \citep{CMasher}.

\section*{Data Availability Statement}

Binned data for making easy comparisons is available in \texttt{ascii} formats at \url{https://github.com/stephenmwilkins/flares_frontier_data}. Data from the wider \flares\ project is available at \url{https://flaresimulations.github.io/data.html}. If you use data from this paper please also cite \citet{FLARES-I} and \citet{FLARES-II}.



\bibliographystyle{mnras}
\bibliography{FLARES-V} 





\bsp	
\label{lastpage}
\end{document}